\documentclass[twocolumn,aps,floats,letterpaper,floatfix,groupedaddress]{revtex4-2}
\usepackage{amsmath}
\usepackage{graphicx}
\usepackage{amsfonts}
\usepackage{amssymb}
\usepackage{bm}
\usepackage{dsfont}
\usepackage{epsfig,float,afterpage}
\usepackage[colorlinks,linkcolor=blue,citecolor=blue,urlcolor=blue]{hyperref}
\usepackage[usenames,dvipsnames]{color}
\newcommand{\beq}{\begin{equation}}
\newcommand{\eeq}{\end{equation}}
\newcommand{\bes}{\begin{subequations}}
\newcommand{\ees}{\end{subequations}}
\newcommand{\bea}{\begin{eqnarray}}
\newcommand{\eea}{\end{eqnarray}}
\newcommand{\ba}{\begin{array}}
\newcommand{\ea}{\end{array}}
\newcommand{\beqn}{\begin{eqnarray*}}
\newcommand{\eeqn}{\end{eqnarray*}}

\newcommand{\f}[2]{\frac{#1}{#2}}

\newcommand{\dg}{\dagger}

\def\nn{\nonumber}

\begin{document}
\title{Quantum noise induced nonreciprocity for single photon transport in parity-time symmetric systems}
\author{Dibyendu Roy$^1$ and G. S. Agarwal$^{2,3,4}$}
\affiliation{$^1$Raman Research Institute, Bangalore 560080, India}
\affiliation{$^2$Institute for Quantum Science and Engineering, Texas A$\&$M University, College Station, TX 77843, USA}
\affiliation{$^3$Department of Physics and Astronomy, Texas A$\&$M University, College Station, TX 77843, USA}
\affiliation{$^4$Department of Biological and Agricultural Engineering, Texas A$\&$M University, College Station, TX 77843, USA}
\begin{abstract}
We show nonreciprocal light propagation for single-photon inputs due to quantum noise in coupled optical systems with gain and loss. We consider two parity-time ($\mathcal{PT}$) symmetric linear optical systems consisting of either two directly coupled resonators or two finite-length waveguides evanescently coupled in parallel. One resonator or waveguide is filled with an active gain medium and the other with a passive loss medium. The light propagation is reciprocal in such $\mathcal{PT}$ symmetric linear systems without quantum noise. We show here that light transmission becomes nonreciprocal when we include quantum noises in our modeling, which is essential for a proper physical description. The quantum nonreciprocity is especially pronounced in the $\mathcal{PT}$ broken phase. Transmitted light intensity in the waveguide of incidence is asymmetric for two waveguides, even without noise. Quantum noise significantly enhances such asymmetry in the broken phase. 
\end{abstract}

\maketitle
\section{Introduction}
Nonreciprocal light propagation through nanoscale optical devices has attracted much theoretical and experimental interests in recent years \cite{Yu2009,RoyPRB2010,Ramezani2010,BiNat2011,KamalNat2011,Fan12,Ramezani12,RoyNatS2013,SounasNatC2013,Bender2013, Peng2014, Chang2014,FratiniPRL2014,LiuPRA2014,EstepNat2014,Yu15,Shi2015,SayrinPRX2015,FratiniPRA2016,RoyPRA2017,Hamann2018,Sohn2021,BiehsPRL2023,BiehsPRB2023, BiehsPRA2023,Upadhyay2024}. One of the most common mechanisms for directional light transmission in optical isolators or diodes is based on magneto-optic Faraday rotation of light employing magnetic fields along light propagation in a magnetically active medium. The other highly explored magnetless mechanisms of nonreciprocal light transmission are different variations of spatio-temporal modulation in linear medium with some momentum conservation rule or momentum bias \cite{Yu2009,KamalNat2011,EstepNat2014,Shi2015,Sohn2021,BiehsPRL2023}, and a combination of Kerr or Kerr-like nonlinearity with space-inversion symmetry (parity) breaking \cite{RoyPRB2010,Fan12,FratiniPRL2014,Yu15,FratiniPRA2016,Hamann2018,Upadhyay2024}. Here, we propose a new mechanism of nonreciprocal light propagation at single-photon level due to quantum noise in parity-time ($\mathcal{PT}$) reversal symmetric linear systems with loss and gain. The presence of loss and gain breaks the time reversal symmetry of the system. Nevertheless, it is well-established now that the sole presence of loss and gain in a linear system is not enough to induce nonreciprocity in light propagation \cite{Ramezani2010, Jalas2013, LiuPRA2014, Peng2014, Chang2014}. One needs either nonlinearities \cite{Ramezani2010, LiuPRA2014, Bender2013, Peng2014, Chang2014} or magneto-optical layer sandwiched between two judiciously balanced gain and loss layers \cite{Ramezani12} to induce optical isolation in such medium with loss and gain. We show that including quantum fluctuations of the fields that are inherent for a medium with gain or loss leads to nonreciprocity in a linear system without a natural or synthetic magnetic field.  More specifically, the quantum noise from the gain medium is enough to induce nonreciprocal light transmission at the single-photon level at zero temperature. Nevertheless, the loss medium is vital in acquiring non-Hermitian $\mathcal{PT}$ symmetry or steady-state transport in such devices. Thus, we need a combination of gain and loss along with intrinsic quantum noises to obtain sizeable nonreciprocity in light transmission without any frequency shift in the output signal at a linear response regime of functionality. We present our results in both the transient and the steady-state domain, the earlier being critical in the $\mathcal{PT}$ broken phase, where the nonreciprocity proliferates with time.    

\begin{figure}
\includegraphics[width=0.7\linewidth]{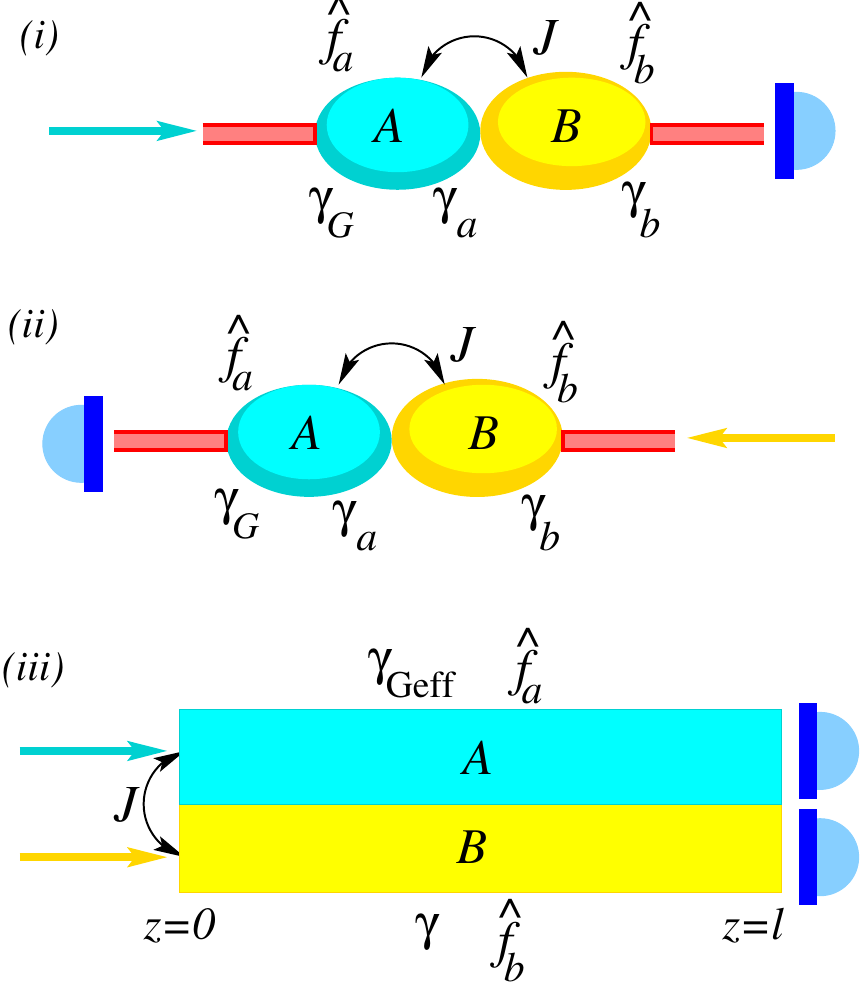}
\caption{Two parity-time $(\mathcal{PT})$ symmetric linear optical systems with active gain and passive loss.  $(i, ii)$ A system composed of two directly coupled resonators connected to two optical fibers at two ends and an incoming single photon from the system's left $(i)$ and right $(ii)$. Resonator $A$ has both gain $(\gamma_G)$ and loss $(\gamma_a)$, and resonator $B$ has only loss $(\gamma_b)$. Gain and loss in resonators lead to quantum fluctuations of the fields represented by noise $\hat{f}_a(t),\hat{f}_b(t)$. $(iii)$ A system consisting of two waveguides of length $l$ next to each other and coupled evanescently with strength $J$. Waveguide $A$ and $B$ are filled, respectively, with effective gain $(\gamma_{\rm Geff})$ and lossy $(\gamma)$ media.  Quantum noise in waveguide $A~ (B)$ is $\hat{f}_a(z)~(\hat{f}_b(z))$. Single-photon inputs at $z=0$ and detectors measuring output intensities at $z=l$ are also shown. }
\label{PTmedium1}
\end{figure}

We consider two $\mathcal{PT}$ symmetric linear optical systems with active gain and passive loss to demonstrate the proposed nonreciprocity due to quantum noise. The first system consists of two resonators or cavities, which are directly coupled to each other in a series configuration (end-to-end connection to form a single path for light propagation), and these resonators are connected to two optical fibers at two ends. Another system comprises two single-mode waveguides evanescently coupled in parallel, generating multiple paths for light propagation. Our coupled systems are depicted in Fig.~\ref{PTmedium1} for incoming light from either side of the coupled resonators or any waveguide. Both these optical systems have been experimentally realized \cite{Ruter2010, Peng2014, Chang2014}. The light propagation in these two systems are a bit similar as we show below. The rest of the paper is organized into three sections for two different coupled systems, including a summary and two appendices for details on calculation.  

\section{Coupled resonators}
First, we explain nonreciprocal light propagation for single-photon inputs in two directly coupled resonators induced by quantum noise. Such a system has been extensively explored for non-Hermitian $\mathcal{PT}$ symmetry and large nonreciprocity in light propagation due to gain-saturated nonlinearity \cite{Chang2014, Peng2014}. Nevertheless, the light transmission in the linear regime of the coupled system is reciprocal regardless of whether the $\mathcal{PT}$ symmetry is broken or unbroken \cite{Peng2014}. We show that the inclusion of quantum noises, which are essential for the correct modeling of the medium, leads to nonreciprocal light propagation for single-photon inputs in the linear regime of this optical system. The left and right resonators have losses, which are the sum of their intrinsic decay rates $\gamma_a,\gamma_b$ and the coupling losses $\kappa_a,\kappa_b$ due to connection to the optical fibers. The left resonator also has a gain of strength $\gamma_G$ due to optical pumping. Thus, our linear system consists of two directly coupled active-passive resonators (e.g., microtoroids in \cite{Chang2014}). Since we are interested in studying light propagation for single-photon inputs, we discuss optical-field dynamics in terms of quantized fields in the resonators. We take $\hat{a}(t),\hat{b}(t)~(\hat{a}^{\dg}(t),\hat{b}^{\dg}(t))$ as the annihilation (creation) operators for the quantum light fields at time $t$ in the left and right resonator, respectively. The gain or loss medium leads to quantum fluctuations of the fields \cite{AgarwalPRA2012}, which are also important to retain the commutation relations of the time-evolved photon-field operators. 
The quantum Langevin equations describing the time-evolution of photon fields in the coupled resonators are \cite{AgarwalPRA2012}:
\bea
&&\frac{d}{dt}\begin{pmatrix}{\hat{a}} \\{\hat{b}}\end{pmatrix}=\mathcal{M}_r\begin{pmatrix}{\hat{a}} \\{\hat{b}}\end{pmatrix}+\begin{pmatrix}{\sqrt{\kappa_a}\:\hat{a}_{\rm in}} \\{\sqrt{\kappa_b}\:\hat{b}_{\rm in}}\end{pmatrix}+\begin{pmatrix}{\hat{f}_a} \\{\hat{f}_b}\end{pmatrix},\label{ev}\\
&&\mathcal{M}_r=\begin{pmatrix}{\f{\gamma_G-\gamma_a-\kappa_a}{2}-i\delta_a} &{iJ}\\{iJ}&{-\f{\gamma_b+\kappa_b}{2}-i\delta_b}\end{pmatrix}, \label{evM}
\eea
where $\hat{a}_{\rm in},\hat{b}_{\rm in}$ are the input fields acting on the left and right resonators. Here, $\delta_a,\delta_b$ are the detuning of the left and right resonator's frequency from the driving frequency. The coupling strength between the left and right resonator fields is $J$. The quantum mechanical fluctuating forces or noises $\hat{f}_a(t)$ and $\hat{f}_b(t)$, respectively, in the left and right resonator are Gaussian variables with zero means, and satisfy the following delta-correlations in time \cite{scully1997quantum,AgarwalPRA2012}:
\bea
&&\langle \hat{f}_a^{\dg}(t)\hat{f}_a(t')\rangle=(\gamma_G+\gamma_an_{\rm th})\delta(t-t'), \nn\\ 
&&\langle \hat{f}_b^{\dg}(t)\hat{f}_b(t')\rangle=\gamma_bn_{\rm th}\delta(t-t'), \nn\\&& \langle \hat{f}_i(t)\hat{f}_j^{\dg}(t')\rangle=\delta_{ij}\gamma_i(n_{\rm th}+1)\delta(t-t'), \label{fd2}
\eea
where $i,j=a,b$ and $n_{\rm th}$ is thermal photon occupancy. These quantum noises are intrinsic to any loss or gain. We are not imposing any external source of noise. Since the frequencies of the resonators' fields are in the optical and infrared range (e.g., 193 THz) in many of these experiments \cite{Peng2014, Chang2014}, $n_{\rm th}$ even at room temperature is of order $10^{-13}$, which safely allows us to drop $n_{\rm th}$ from the noise correlations in Eq.~\ref{fd2} for rest of the calculations below. Nevertheless, it might be necessary to retain $n_{\rm th}$ in Eq.~\ref{fd2} for other frequencies (e.g., microwave) of the resonators' fields. The expectations $\langle \dots \rangle$ in Eq.~\ref{fd2} are over the realization of noises. Here, we note a difference in the nature of correlations between quantum noises in the active (amplifying) and passive (absorbing) resonator. These quantum white noises would contribute to the light transmission in the coupled resonators and significantly modify the nature of light transmission, as discussed below.


We clarify that we do not work with $\mathcal{PT}$-symmetric quantum mechanics as this is flawed \cite{Lee2014,Tang2016}. All the difficulties arise from using non-Hermitian Hamiltonian literally in dynamical evolution. In all optical systems so far studied both experimentally \cite{Ruter2010, Peng2014, Chang2014} or theoretically, the $\mathcal{PT}$ symmetry is used at the level of the mean value equations, e.g., Eq.~\ref{ev} for $\langle \hat{a}\rangle$ and $\langle \hat{b}\rangle$ without the noise terms. These experiments do not realize $\mathcal{PT}$-symmetric quantum mechanics. Our full quantum mechanical equations are in the framework of open quantum system theory as, for example, developed in the context of lasers, and the theoretical description of quantum noise follows what is described in standard laser texts see, for example, \cite{scully1997quantum}.

We are interested in finding the transmitted light intensities through the coupled resonators for an input field either from the left or the right of the coupled system, e.g., either $\hat{a}_{\rm in} \ne 0, \hat{b}_{\rm in}=0$ or  $\hat{a}_{\rm in}=0, \hat{b}_{\rm in}\ne 0$. The relations between the input and output fields are: $\hat{a}_{\rm out}+\hat{a}_{\rm in}=\sqrt{\kappa_a}\hat{a}(t)$ and $\hat{b}_{\rm out}+\hat{b}_{\rm in}=\sqrt{\kappa_b}\hat{b}(t)$, where $\hat{a}_{\rm out}$ and $\hat{b}_{\rm out}$ are the output fields \cite{Walls2007, RoyPRA2017, RoyRMP2017}. Hereafter, we take $\langle \hat{a}^{\dg}_{\rm in}\hat{a}_{\rm in}\rangle=\langle \hat{b}^{\dg}_{\rm in}\hat{b}_{\rm in}\rangle=\mathcal{I}_{\rm in}$ for a single-photon input field from the left or right of the system. We first set $\gamma_a+\kappa_a=\gamma_b+\kappa_b=\gamma$, and $\delta_a=\delta_b=0$ for simplicity. For the $\mathcal{PT}$ symmetry of the coupled resonators, a balance gain and loss requires $\gamma_G=2\gamma$, which forbids the coupled resonators from attaining a steady state. So, we now explore transient light propagation in the $\mathcal{PT}$ symmetry unbroken $(\gamma<2J)$ and broken $(\gamma>2J)$ phase of the model.

\begin{figure}
\includegraphics[width=\linewidth]{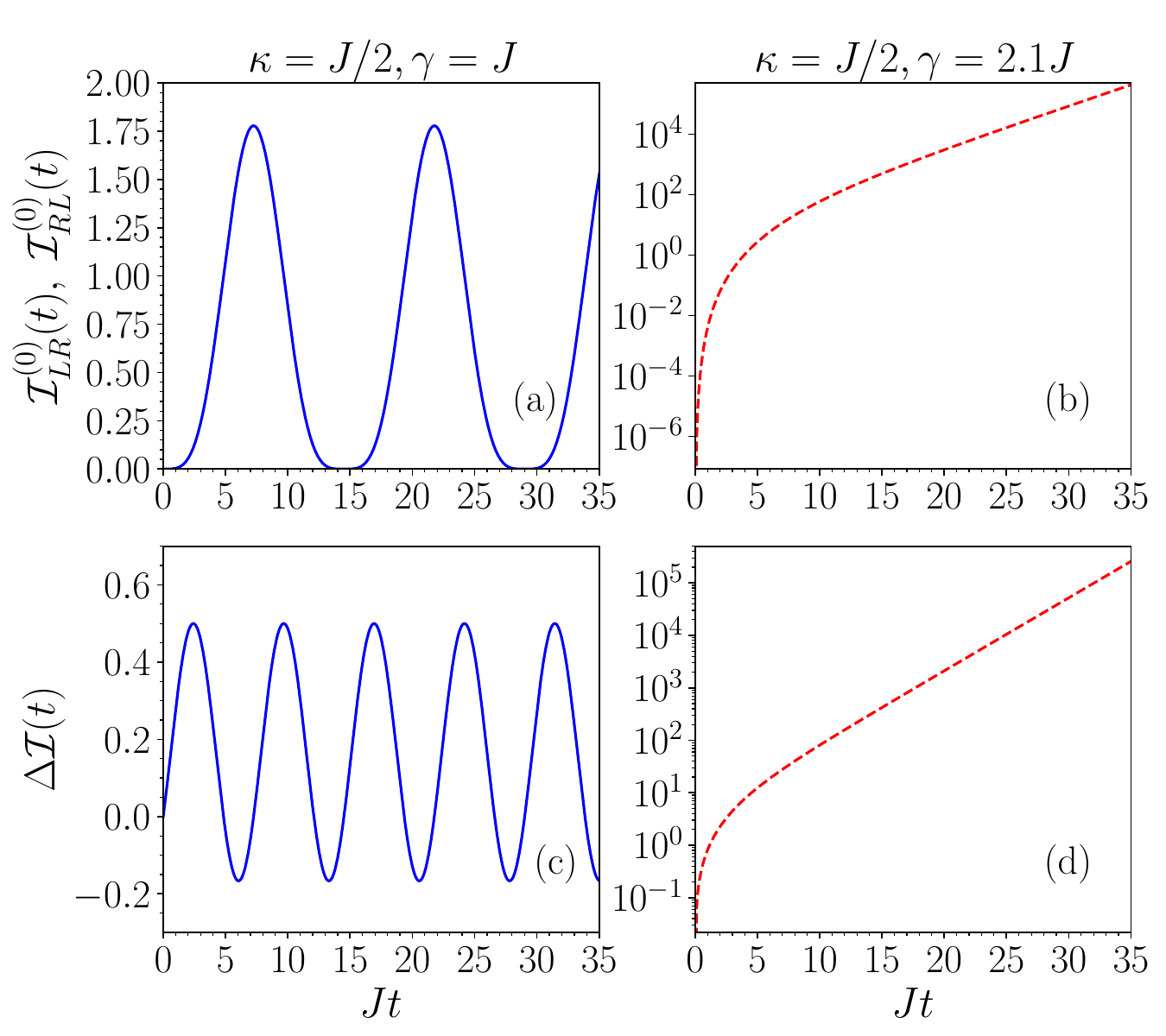}
\caption{Time evolution of reciprocal transmitted output intensity without noise $(\mathcal{I}^{(0)}_{LR}(t),\mathcal{I}^{(0)}_{RL}(t))$ and nonreciprocity due to quantum noise $(\Delta \mathcal{I}(t))$ in $\mathcal{PT}$ symmetric coupled resonators in the unbroken (first column) and broken phase (second column).   }
\label{PTsymCR}
\end{figure}

Let us define $\mathcal{I}_{LR}(t)~(\mathcal{I}_{RL}(t))$ as the time-dependent transmitted light intensity from the left to right (right to left) of the system for a single-photon input field from the left (right). These intensities can be found by applying the above input-output relations. For example, we derive $\mathcal{I}_{LR}(t)=\langle \hat{b}_{\rm out}^{\dg}\hat{b}_{\rm out}\rangle=\kappa_b\langle \hat{b}^{\dg}(t)\hat{b}(t) \rangle$ by finding time evolution of the fields in Eq.~\ref{ev} and using the noise properties in Eq.~\ref{fd2}. These intensities can further be expressed as $\mathcal{I}_{LR}(t)=\mathcal{I}^{(0)}_{LR}(t)+\mathcal{I}^{(n)}_{L R}(t)$ and $\mathcal{I}_{R L}(t)=\mathcal{I}^{(0)}_{R L}(t)+\mathcal{I}^{(n)}_{R L}(t)$ by separating the contribution in the absence of noise $(\mathcal{I}^{(0)}_{LR}(t),\mathcal{I}^{(0)}_{RL}(t))$ and that due to the noise (see App.~\ref{TD} for derivation). In the absence of quantum noise, we find a reciprocal time-dependent transmitted intensity in the $\mathcal{PT}$ symmetric system, which are 
\bea
\mathcal{I}^{(0)}_{LR}(t)=\mathcal{I}^{(0)}_{R L}(t)=\frac{64J^2\kappa^2\mathcal{I}_{\rm in}}{(4J^2-\gamma^2)^2}\sin^4(\frac{t}{4}\sqrt{4J^2-\gamma^2}),
\eea
where we take $\kappa_a=\kappa_b=\kappa$. This result of reciprocal light propagation agrees with the previous studies for such systems in the linear regime without quantum noises \cite{Ramezani2010, Jalas2013, LiuPRA2014, Peng2014, Chang2014}. These transmitted output intensities oscillate with time in the unbroken phase and grow rapidly with time in the broken phase. We depict these features in Figs.~\ref{PTsymCR}(a,b) for $\mathcal{I}_{\rm in}=1$. The noise contributions $\mathcal{I}^{(n)}_{LR}(t)$ and $\mathcal{I}^{(n)}_{R L}(t)$ for the $\mathcal{PT}$ symmetric system at time $t$ are
\bea
\mathcal{I}^{(n)}_{R L}(t)&=\frac{2\kappa\gamma^2}{4J^2-\gamma^2}\Big(1+2J^2t/\gamma-\cos\big(t\sqrt{4J^2-\gamma^2}\:\big)\Big)\nn\\&+\frac{2\kappa\gamma(2J^2-\gamma^2)}{(4J^2-\gamma^2)^{3/2}}\sin\big(t\sqrt{4J^2-\gamma^2}\:\big),\label{tna}\\
\mathcal{I}^{(n)}_{L R}(t)&=\frac{4\kappa \gamma J^2}{4J^2-\gamma^2}\big(t-\frac{1}{\sqrt{4J^2-\gamma^2}}\sin\big(t\sqrt{4J^2-\gamma^2}\:\big)\big).\label{tnb}
\eea
While $\mathcal{I}^{(0)}_{LR}(t)$ and $\mathcal{I}^{(0)}_{R L}(t)$ are proportional to input photon number $(\mathcal{I}_{\rm in})$, the noise contributions $\mathcal{I}^{(n)}_{LR}(t)$ and $\mathcal{I}^{(n)}_{R L}(t)$ are independent of the photon number. Since $\mathcal{I}^{(n)}_{L R}(t) \ne \mathcal{I}^{(n)}_{R L}(t)$ from Eqs.~\ref{tna},\ref{tnb}, we have $\mathcal{I}_{L R}(t) \ne \mathcal{I}_{R L}(t)$. Thus, the time-dependent transmitted intensity in the $\mathcal{PT}$ symmetric coupled resonators is nonreciprocal due to quantum noise from the active (amplifying) resonator. The nonreciprocity will be most pronounced for incident beams at single photon level. The nonreciprocity in the transmitted intensity is 
\begin{align}
\Delta \mathcal{I}(t)=&\mathcal{I}^{(n)}_{RL}-\mathcal{I}^{(n)}_{L R}=\frac{4\kappa\gamma^2}{4J^2-\gamma^2}\sin^2\big(\frac{t}{2}\sqrt{4J^2-\gamma^2}\big)\nn\\&+\frac{2\kappa\gamma}{\sqrt{4J^2-\gamma^2}}\sin(t\sqrt{4J^2-\gamma^2}),
\end{align}
which oscillates with $t$ in the unbroken phase as we depict in Fig.~\ref{PTsymCR}(c). The nonreciprocity in the broken phase grows rapidly with time as shown in Fig.~\ref{PTsymCR}(d). Therefore, the nonreciprocity in broken phase would be useful for potential application of optical isolators. One physically relevant quantity of interest is the relative nonreciprocity, defined as a ratio between $\Delta \mathcal{I}(t)$ and the reciprocally transmitted intensity without quantum noise. The relative nonreciprocity $\Delta \mathcal{I}(t)/\mathcal{I}^{(0)}_{LR}(t)$ diverges in the unbroken phase when $\mathcal{I}^{(0)}_{LR}(t)$ vanishes at certain time intervals. In the broken phase, the relative nonreciprocity saturates at long time $(t>2/\sqrt{\gamma^2-4J^2})$ to $\gamma(\gamma^2-4J^2)(\gamma+\sqrt{\gamma^2-4J^2})/(4J^2\kappa \mathcal{I}_{\rm in})$, which can become large as we move deeper inside the broken phase.

We notice above that the $\mathcal{PT}$ symmetry of the model is achieved for restricted values of $\gamma$ and $\gamma_G$, which deny the steady state of Eq.~\ref{ev}. The steady state is reached when the eigenvalues $\Lambda_{\pm}=(\gamma_G-2\gamma \pm \sqrt{\gamma_G^2-16J^2})/4$ of $\mathcal{M}_r$ satisfy Re$[\Lambda_{\pm}]<0$, which can be obtained for either (i) $J>0$ when $\gamma_G \le \gamma$ or (ii) $J>\sqrt{\gamma(\gamma_G-\gamma)}/2$ when $\gamma<\gamma_G<2\gamma$. We denote the steady-state transmitted intensity from the left to right (right to left) of the coupled resonators for a single-photon input by $\mathcal{I}_{LR}~(\mathcal{I}_{RL})$. 
We find for an incoming single photon from the left of the coupled system (see App.~\ref{TID} for derivation):
\begin{align}
\mathcal{I}_{LR}&=\lim_{t \to \infty}\kappa_b\langle \hat{b}^{\dg}(t)\hat{b}(t) \rangle=\frac{16J^2\kappa_a \kappa_b\mathcal{I}_{\rm in}}{|4J^2+\gamma(\gamma-\gamma_G)|^2}\nn\\&+\frac{4\gamma_G \kappa_bJ^2}{(2\gamma-\gamma_G)(4J^2+\gamma(\gamma-\gamma_G))},\label{transLRsp}
\end{align}
where the first term in the right hand side of the above expression appears even in the absence of the noise terms in Eq.~\ref{ev}, and the second term emerges solely due to the quantum noise $\hat{f}_a(t)$ in the left resonator. A similar calculation for a single-photon driving from the right of the system gives 
\begin{align}
\mathcal{I}_{RL}&=\lim_{t \to \infty}\kappa_a\langle \hat{a}^{\dg}(t)\hat{a}(t) \rangle=\frac{16J^2\kappa_a\kappa_b \mathcal{I}_{\rm in}}{|4J^2+\gamma(\gamma-\gamma_G)|^2}\nn\\
&+\frac{\gamma_G\kappa_a(4J^2+\gamma(2\gamma-\gamma_G))}{(2\gamma-\gamma_G)(4J^2+\gamma(\gamma-\gamma_G))},\label{transRLsp}
\end{align}
where again the first term in the right-hand side of the above expression is present even without the noise terms in Eq.~\ref{ev}, and the second term emerges solely due to the quantum noise $\hat{f}_a(t)$ in the active resonator. Thus, we find from Eqs.~\ref{transLRsp},\ref{transRLsp} that the steady-state transmission of a single input photon is reciprocal in the absence of quantum noise $\hat{f}_a(t)$ since the first terms in Eqs.~\ref{transLRsp},\ref{transRLsp} are the same. We further observe from Eqs.~\ref{transLRsp},\ref{transRLsp} that $\mathcal{I}_{RL}>\mathcal{I}_{LR}$ in the steady state, and the difference is $\Delta \mathcal{I} \equiv \mathcal{I}_{RL}-\mathcal{I}_{LR}=\gamma\gamma_G\kappa/(4J^2+\gamma(\gamma-\gamma_G))$, where we again take $\kappa_a=\kappa_b=\kappa$.  Therefore, the steady-state light propagation is also nonreciprocal due to quantum noise. 

The steady-state relative nonreciprocity is $(4J^2+\gamma(\gamma-\gamma_G))\gamma\gamma_G/(16J^2\kappa\mathcal{I}_{\rm in})$. The relative nonreciprocity saturates to $\gamma\gamma_G/(4\kappa\mathcal{I}_{\rm in})$ at large $J$ when the reciprocal transmission is also significant. The saturation value of relative nonreciprocity can be large for single-photon inputs $(\mathcal{I}_{\rm in} \approx 1)$ since $\kappa<\gamma$ and $\gamma_G$ can be chosen near $2\gamma$. For weak coherent state inputs, the relative nonreciprocity falls with increasing amplitude of the coherent incoming light as the value of quantum noise-induced nonreciprocity remains constant, and the reciprocal light transmission grows with the amplitude of the coherent state.

\section{Coupled waveguides}
The light propagation in two directly coupled resonators is a bit similar to that in another well-explored $\mathcal{PT}$ symmetric linear system of coupled waveguides in Fig.~\ref{PTmedium1}$(iii)$. The waveguide $A$ is filled with a gain medium, which is optically pumped by an external source to provide a gain coefficient $\gamma_G$ for guided light in the waveguide. The parallel waveguides $A$ and $B$ dissipate energy with a loss coefficient $\gamma$ to external environments. Thus, the effective gain coefficient in the waveguide $A$ is $\gamma_{\rm Geff}=\gamma_G-\gamma$. 
We take $\hat{a}(z),\hat{b}(z)~(\hat{a}^{\dg}(z),\hat{b}^{\dg}(z))$ as the annihilation (creation) operators for the quantum light fields at position $z$ of the waveguide $A$ and $B$, respectively. The single-photon input states are $|1,0\rangle \equiv \hat{a}^{\dg}(0)|\varphi\rangle$ and $|0,1\rangle \equiv \hat{b}^{\dg}(0)|\varphi\rangle$ for an incident photon in the waveguide $A$ and $B$, respectively. Here, $|\varphi\rangle$ denotes the vacuum of electromagnetic fields. 
Again, the gain or loss leads to quantum fluctuations in the fields \cite{AgarwalPRA2012}. The quantum Langevin equations describing the spatial evolution of photon-field operators in the coupled-waveguide system are \cite{AgarwalPRA2012}:
\begin{align}
  \frac{d}{dz}\begin{pmatrix}{\hat{a}} \\{\hat{b}}\end{pmatrix}=\mathcal{M}_w\begin{pmatrix}{\hat{a}} \\{\hat{b}}\end{pmatrix}+\begin{pmatrix}{\hat{f}_a} \\{\hat{f}_b}\end{pmatrix},~\mathcal{M}_w=\begin{pmatrix}{\frac{\gamma_{\rm Geff}}{2}} &{iJ}\\{iJ}&{-\frac{\gamma}{2}}\end{pmatrix},\label{qf1}
\end{align}
for $0<z<l$. The coupling strength between two quantized fields in the waveguide $A$ and $B$ is $J$. The eigenvalues of $\mathcal{M}_w$ are the same as $\Lambda_{\pm}$, and the coupled-waveguide system has $\mathcal{PT}$ symmetry when $\gamma_{\rm Geff}=\gamma$ or $\gamma_G=2\gamma$. The non-Hermitian system described by $\mathcal{M}_w$ undergoes a $\mathcal{PT}$ symmetry unbroken to broken phase transition  at $\gamma/(2J)=1$ as $\Lambda_{\pm}$ switch from imaginary for $\gamma<2J$ to real for $\gamma>2J$. The quantum noises $\hat{f}_a(z)$ and $\hat{f}_b(z)$, respectively, in the waveguide $A$ and $B$ are again chosen as Gaussian variables with zero means, and they satisfy the delta-correlations in position at zero temperature. These noise correlations are the same as those in Eq.~\ref{fd2} after replacing $t,t'$ by $z,z'$ and taking $\gamma_a=\gamma_b=\gamma$. We again take $n_{\rm th}\to 0$ here.

The transmission of an incident photon from $z=0$ of the waveguide $A$ to $z=l$ of the waveguide $B$ and $A$ are related, respectively, to the transmission and reflection of an incident photon from the left of the coupled resonators (see App.~\ref{WG}). Let us define $\mathcal{I}^{(0)}_{\alpha\beta}$ for $\alpha,\beta=A,B$ as the output intensity of a single-photon input from $z=0$ of the waveguide $\alpha$ to $z=l$ of the waveguide $\beta$ in the absence of noises. For example, $\mathcal{I}^{(0)}_{AB}=\langle 1,0|\hat{b}^{\dg}(l)\hat{b}(l)|1,0\rangle$. We find for the $\mathcal{PT}$ symmetric coupled waveguides (see App.~\ref{WG}):
\begin{align}
&\mathcal{I}^{(0)}_{AB}=\mathcal{I}^{(0)}_{BA}=\frac{4J^2 \sinh^2(l\sqrt{\gamma^2-4J^2}/2)}{\gamma^2-4J^2},\label{transabPT} \\
&\mathcal{I}^{(0)}_{AA,BB}=\Big|\cosh\big(\frac{l}{2}\sqrt{\gamma^2-4J^2}\big) \pm \frac{\gamma\sinh\big(\frac{l}{2}\sqrt{\gamma^2-4J^2}\big)}{\sqrt{\gamma^2-4J^2}}\Big|^2.\label{transAAPT}
\end{align}
Eq.~\ref{transabPT} infers that the light transmission from waveguide $A$ to $B$ is the same as that from waveguide $B$ to $A$ without quantum noises, which is similar to the coupled resonators. $\mathcal{I}^{(0)}_{\alpha\beta}$ in Eqs.~\ref{transabPT}, \ref{transAAPT} oscillates with length $l$ of the waveguides with a finite amplitude of $4J^2/(4J^2-\gamma^2)$, when the optical system is in the $\mathcal{PT}$ symmetry unbroken phase for $\gamma<2J$ (see Fig.~\ref{PTsymCW} of App.~\ref{WG}). However, these output intensities exponentially diverge with increasing $l$ in the $\mathcal{PT}$ symmetry broken phase for $\gamma>2J$. Then, we have the following asymptotic forms:
\bea
&&\mathcal{I}^{(0)}_{AA,BB} \to \frac{\gamma^2-2J^2\pm\gamma\sqrt{\gamma^2-4J^2}}{2(\gamma^2-4J^2)}e^{l\sqrt{\gamma^2-4J^2}},\nn\\
&&\mathcal{I}^{(0)}_{AB}=\mathcal{I}^{(0)}_{BA} \to \frac{J^2}{\gamma^2-4J^2}e^{l\sqrt{\gamma^2-4J^2}},
\eea
which predict $\mathcal{I}^{(0)}_{AA}>\mathcal{I}^{(0)}_{BA},\mathcal{I}^{(0)}_{AB}>\mathcal{I}^{(0)}_{BB}$ beyond a critical length. In the limit $\gamma \to 2J$, we further find $\mathcal{I}^{(0)}_{AB}=\mathcal{I}^{(0)}_{BA}=(Jl)^2, \mathcal{I}^{(0)}_{AA}=(Jl+1)^2, \mathcal{I}^{(0)}_{BB}=(Jl-1)^2$, which infer $\mathcal{I}^{(0)}_{AA}>\mathcal{I}^{(0)}_{BA},\mathcal{I}^{(0)}_{AB}>\mathcal{I}^{(0)}_{BB}$ when $l>1/(2J)$. In the $\mathcal{PT}$ broken phase beyond a critical $l$, the output intensities in the waveguide $A$ are therefore higher than that in the waveguide $B$ for an input from either waveguide. This seems to match with the results of \textcite{Ruter2010} for a classical input light without quantum noise. $\mathcal{I}^{(0)}_{AA}>\mathcal{I}^{(0)}_{BB}$ also implies an asymmetric reflection in the coupled resonators \cite{Regensburger2012, Feng2013} when we apply the above analogy in light propagation between the coupled waveguides and resonators (see App.~\ref{Coures} for details).   

We next discuss output intensities $\mathcal{I}_{\alpha\beta}$ of a single-photon input at $z=0$ of the waveguide $\alpha$ to $z=l$ of the waveguide $\beta$ in the presence of noises. These are found after averaging the output intensities over the noises. We find $\mathcal{I}_{AA}=\mathcal{I}^{(0)}_{AA}+\mathcal{I}^{(n)}_{A}, \mathcal{I}_{AB}=\mathcal{I}^{(0)}_{AB}+\mathcal{I}^{(n)}_{B}, \mathcal{I}_{BA}=\mathcal{I}^{(0)}_{BA}+\mathcal{I}^{(n)}_{A}$, and $\mathcal{I}_{BB}=\mathcal{I}^{(0)}_{BB}+\mathcal{I}^{(n)}_{B}$, where $\mathcal{I}^{(n)}_{A}$ and $\mathcal{I}^{(n)}_{B}$ are the contributions of the gain medium's quantum noise to the output light intensities, respectively, in the waveguide $A$ and $B$. Interestingly, the noise contributions $\mathcal{I}^{(n)}_B$ and $\mathcal{I}^{(n)}_A$ in the $\mathcal{PT}$ symmetric system are identical to $\mathcal{I}^{(n)}_{LR}(t)$ and $\mathcal{I}^{(n)}_{RL}(t)$, respectively, when we replace $t$ by $l$ and set $\kappa=1$. 
Since $\mathcal{I}^{(n)}_{A} \ne \mathcal{I}^{(n)}_{B}$ both in the unbroken and broken phase, we get $\mathcal{I}_{AB} \ne \mathcal{I}_{BA}$ leading to nonreciprocal light propagation between the waveguide $A$ to $B$ and the waveguide $B$ to $A$ due to inclusion of noise in the system. The nonreciprocity $\Delta \mathcal{I}(l)=\mathcal{I}^{(n)}_{A}-\mathcal{I}^{(n)}_{B}$ oscillates with $l$ in the unbroken phase, and it grows with an increasing $l$ in the broken phase (see Fig.~\ref{PTsymCW} of App.~\ref{WG}). 

\section{Summary and outlook}
In summary, we have shown the emergence of nonreciprocity in $\mathcal{PT}$ symmetric linear optical systems at the single-photon level due to including quantum noises in a rigorous, mathematically consistent modeling of the gain and loss in such systems \cite{AgarwalPRA2012}. Such nonreciprocity stems from the spontaneous generation of photons, which leads to the non-classical behavior of light fields, including strongly correlated eigenmodes of the systems. As explained above, our modeling of $\mathcal{PT}$ symmetric linear optical systems with quantum noise correctly describes two experimentally realized models of coupled waveguides and resonators. 
We must generalize the above description of quantum noises from coupled resonators or waveguides to extended systems such as synthetic photonic lattices, which were explored for unidirectional or asymmetric reflection in $\mathcal{PT}$ symmetric metamaterials \cite{Regensburger2012, Feng2013}. The $\mathcal{PT}$ symmetric lattices, e.g., $\mathcal{PT}$ symmetric Su-Schrieffer-Heeger chains, can also host topological properties, including zero modes \cite{Qian2024}, and it would be exciting to examine how the inclusion of quantum noises affects their topological and transport features. Our description of transport in an active medium with quantum noises applies to mechanical, opto-mechanical, and electrical systems beyond visible frequencies. Finally, it would also be helpful to extend the current description, including quantum noises, to nonlinear $\mathcal{PT}$ systems \cite{Ramezani2010, LiuPRA2014, RoyPRA2017} to verify the interplay of nonlinearity and quantum noises for nonreciprocity.

\section{Acknowledgments}
DR thanks Rupak Bag for discussion. GSA thanks the support of Air Force Office of Scientific Research (Award No FA-9550-20-1-0366) and the Robert A Welch Foundation (A-1943-20210327). GSA also thanks INFOSYS FOUNDATION CHAIR of the IISc, Bangalore, which made this collaboration possible. 

\appendix

\section{Two directly coupled resonators}\label{Coures}
We here outline the general procedure for solving quantum Langevin equations in Eqs.~\ref{ev},\ref{qf1} and then obtaining the physical quantities like output photon numbers. These are the first-order linear differential equations and can be integrated to obtain $\hat{a}(t)$ and $\hat{b}(t)$ (or $\hat{a}(z),\hat{b}(z)$) in terms of the initial conditions and input fields. In this section, we provide the details of the various results for two direct coupled resonators connected to two optical fibers at two ends, and an incoming single photon from the left and right of the system.

\subsection{Time-dependent photon transport in $\mathcal{PT}$ symmetric resonators}\label{TD}
We here take $\delta_1=\delta_2=0$ and $\gamma_G=2\gamma$ for $\mathcal{PT}$ symmetry of the linear system. The calculations are simplified by using canonical diagonalization of the non-Hermitian matrix $\mathcal{M}_r$ 
\bea
 \mathcal{V}\mathcal{M}_r\mathcal{V}^{-1}=\begin{pmatrix}{\lambda_-} &{0}\\{0}&{\lambda_+}\end{pmatrix},~~ \lambda_{\mp}=\mp \frac{1}{2} \sqrt{\gamma^2-4J^2}. \label{UmatCR}
\eea
The eigenvalues $\lambda_{\mp}$ of $\mathcal{M}_r$ are in general complex since $\mathcal{M}_r$ is non-Hermitian. This is especially in the $\mathcal{PT}$ broken phase. The diagonalizing matrix $\mathcal{V}$ is not unitary. It is to be noted that $\mathcal{V}^{-1}$ is constructed out of the right eigenfunctions of $\mathcal{M}_r$, whereas $\mathcal{V}$ is constructed out of the left eigenfunctions of $\mathcal{M}_r$. To integrate Eq.~\ref{ev}, it is convenient to apply the matrix $\mathcal{V}$ to transform the operators $\hat{a}$ and $\hat{b}$ to $\hat{c}_{\pm}$ as
\bea
\begin{pmatrix}{\hat{c}_-} \\{\hat{c}_+}\end{pmatrix}=\mathcal{V}\begin{pmatrix}{\hat{a}} \\{\hat{b}}\end{pmatrix},~~\begin{pmatrix}{\hat{a}} \\{\hat{b}}\end{pmatrix}=\mathcal{V}^{-1}\begin{pmatrix}{\hat{c}_-} \\{\hat{c}_+}\end{pmatrix}\label{NM1}.
\eea
It should be borne in mind that we use $\hat{c}_{\pm}$ as intermediate steps. These do not satisfy bosonic commutation relations as $\mathcal{V}$ is not unitary. It is easily seen that
\bea
\mathcal{V}=\begin{pmatrix}{\f{i(\lambda_0-\gamma)}{2J}} &{1}\\{\f{-i(\lambda_0+\gamma)}{2J}}&{1}\end{pmatrix},\:\mathcal{V}^{-1}=\frac{J}{i\lambda_0}\begin{pmatrix}{1}&{-1}\\{\f{i(\lambda_0+\gamma)}{2J}} &{\f{i(\lambda_0-\gamma)}{2J}}\end{pmatrix},\nn\\
\eea
where $\lambda_0=\sqrt{\gamma^2-4J^2}$ and $\mathcal{V}\mathcal{V}^{-1}=\mathcal{V}^{-1}\mathcal{V}=\mathbf{1}$. The advantage of applying Eq.~\ref{NM1} is that $\hat{c}_{\pm}$ satisfy uncoupled equations:
\bea
\frac{d}{dt}\hat{c}_{\mp}&=&\lambda_{\mp}\hat{c}_{\mp}+ \hat{c}_{{\rm in},\mp}+\hat{f}_{\mp},~{\rm with}~\label{NM2}\\
\begin{pmatrix}{\hat{c}_{{\rm in},-}} \\{\hat{c}_{{\rm in},+}}\end{pmatrix}&=&\mathcal{V}\begin{pmatrix}{\sqrt{\kappa_a}\:\hat{a}_{\rm in}} \\{\sqrt{\kappa_b}\:\hat{b}_{\rm in}}\end{pmatrix},{\rm and}\begin{pmatrix}{\hat{f}_-} \\{\hat{f}_+}\end{pmatrix}=\mathcal{V}\begin{pmatrix}{\hat{f}_a} \\{\hat{f}_b}\end{pmatrix}.
\eea
We can find formal solution for time evolution of these operators in Eq.~\ref{NM1} as
\bea
\hat{c}_{\mp}(t)=e^{\lambda_{\mp}t}\hat{c}_{\mp}(0)+\frac{e^{\lambda_{\mp}t}-1}{\lambda_{\mp}}\hat{c}_{{\rm in},\mp}+\int_{0}^{t}d\tau\:e^{\lambda_{\mp}(t-\tau)}\hat{f}_{\mp}(\tau),\nn\\ \label{SolNM}
\eea
where $\hat{c}_{\mp}(0)$ are the initial condition of these operators, and we set them here zero. We can further find the following correlations using the formal solution of these operators in Eq.~\ref{SolNM} and the noise properties in Eq.~\ref{fd2} with $n_{\rm th}\to 0$. 
\bea
\langle \hat{c}_-^{\dg}(t) \hat{c}_-(t) \rangle &=&|\mathcal{V}_{11}|^2 \Big(\kappa_a \mathcal{I}_{\rm in} \frac{|e^{\lambda_-t}-1|^2}{|\lambda_-|^2}\nn\\&&+\gamma_G \frac{e^{(\lambda_-+\lambda_-^*)t}-1}{\lambda_-+\lambda_-^*}\Big),\label{cor1}\\
\langle \hat{c}_+^{\dg}(t) \hat{c}_+(t) \rangle &=&|\mathcal{V}_{21}|^2 \Big(\kappa_a \mathcal{I}_{\rm in} \frac{|e^{\lambda_+t}-1|^2}{|\lambda_+|^2}\nn\\&&+\gamma_G \frac{e^{(\lambda_++\lambda_+^*)t}-1}{\lambda_++\lambda_+^*}\Big),\label{cor2}\\
\langle \hat{c}_+^{\dg}(t) \hat{c}_-(t) \rangle &=&\mathcal{V}_{11}\mathcal{V}_{21}^*\Big(\kappa_a \mathcal{I}_{\rm in} \frac{(e^{\lambda_-t}-1)(e^{\lambda_+^*t}-1)}{\lambda_-\lambda_+^*}\nn\\&&+\gamma_G \frac{e^{(\lambda_-+\lambda_+^*)t}-1}{\lambda_-+\lambda_+^*}\Big),\label{cor3}\\
\langle \hat{c}_-^{\dg}(t) \hat{c}_+(t) \rangle &=&\mathcal{V}^*_{11}\mathcal{V}_{21}\Big(\kappa_a \mathcal{I}_{\rm in} \frac{(e^{\lambda_-^*t}-1)(e^{\lambda_+t}-1)}{\lambda_-^*\lambda_+}\nn\\&&+\gamma_G \frac{e^{(\lambda_-^*+\lambda_+)t}-1}{\lambda_-^*+\lambda_+}\Big).\label{cor4}
\eea
Here, $\mathcal{V}_{ij}$ are elements of $\mathcal{V}$, and $\langle \dots \rangle$ denotes an expectation in the initial state of the full system and an averaging over the quantum noises.

{\it Input from left:} Let us first consider an input from the left of the coupled resonators, i.e., $\hat{a}_{\rm in} \ne 0$ and $\hat{b}_{\rm in}=0$. The transmitted output intensity in the right optical fiber of the system is
\bea
&&\mathcal{I}_{LR}(t)=\langle \hat{b}^{\dg}_{\rm out}\hat{b}_{\rm out} \rangle\nn\\
&&=\langle (-\hat{b}^{\dg}_{\rm in}+\sqrt{\kappa_b}\:\hat{b}^{\dg}(t))(-\hat{b}_{\rm in}+\sqrt{\kappa_b}\:\hat{b}(t)) \rangle \nn\\
&&= \kappa_b \langle \hat{b}^{\dg}(t) \hat{b}(t)\rangle \nn\\
&&=\kappa_b\big(|\mathcal{V}^{-1}_{22}|^2 \langle \hat{c}_+^{\dg}(t) \hat{c}_+(t) \rangle +|\mathcal{V}^{-1}_{21}|^2 \langle \hat{c}_-^{\dg}(t) \hat{c}_-(t)\rangle\nn\\
&&+\mathcal{V}^{-1*}_{22}\mathcal{V}^{-1}_{21} \langle \hat{c}_+^{\dg}(t) \hat{c}_-(t) \rangle +\mathcal{V}^{-1*}_{21}\mathcal{V}^{-1}_{22} \langle \hat{c}_-^{\dg}(t) \hat{c}_+(t)\rangle \big),\nn\\\label{OILR}
\eea
where we use the transformation in Eq.~\ref{NM1}. Plugging these correlators from Eqs.~\ref{cor1}-\ref{cor4} in Eq.~\ref{OILR}, we  simplify it by using the explicit forms of $\lambda_{\mp}, \mathcal{V}_{ij}$, and $\mathcal{V}^{-1}_{ij}$. We separate $\mathcal{I}_{LR}(t)$ in two parts due to the contribution without $(\mathcal{I}_{LR}^{(0)}(t))$ and with the quantum noise $(\mathcal{I}_{LR}^{(n)}(t))$ as:
\bea
&&\mathcal{I}_{LR}(t)=\mathcal{I}_{LR}^{(0)}(t)+\mathcal{I}_{LR}^{(n)}(t),\nn\\
&&\mathcal{I}_{LR}^{(0)}(t)=\frac{64J^2\kappa_a\kappa_b\mathcal{I}_{\rm in}}{(4J^2-\gamma^2)^2}\sin^4(\frac{t}{4}\sqrt{4J^2-\gamma^2}),\label{IOLR0}\\
&&\mathcal{I}_{LR}^{(n)}(t)=\frac{4\kappa_b \gamma J^2}{4J^2-\gamma^2}\big(t-\frac{1}{\sqrt{4J^2-\gamma^2}}\sin\big(t\sqrt{4J^2-\gamma^2}\:\big)\big).\nn\\\label{IOLRn}
\eea
The expressions in Eqs.~\ref{IOLR0} and \ref{IOLRn} are valid for both in the unbroken and broken phase of the $\mathcal{PT}$ symmetric system. We have used ${\rm lim}_{\lambda \to 0}(e^{\lambda t}-1)/\lambda=t$ in writing the last expression. For an input from the left of the coupled resonators, the reflected output intensity in the left optical fiber is
\bea
&&\mathcal{I}_{LL}(t)=\langle \hat{a}^{\dg}_{\rm out}\hat{a}_{\rm out} \rangle\nn\\
&&=\langle (-\hat{a}^{\dg}_{\rm in}+\sqrt{\kappa_a}\:\hat{a}^{\dg}(t))(-\hat{a}_{\rm in}+\sqrt{\kappa_a}\:\hat{a}(t)) \rangle \nn\\
&&= \mathcal{I}_{\rm in}+ \kappa_a \langle \hat{a}^{\dg}(t) \hat{a}(t)\rangle-\sqrt{\kappa}_a(\langle \hat{a}^{\dg}_{\rm in} \hat{a}(t)\rangle+\langle\hat{a}^{\dg}(t)\hat{a}_{\rm in}\rangle).\nn\\\label{OILL1}
\eea
We again evaluate each parts of the above expression separately. We get
\bea
&&\kappa_a\langle \hat{a}^{\dg}(t) \hat{a}(t)\rangle=\big(|\mathcal{V}^{-1}_{11}|^2 \langle \hat{c}_-^{\dg}(t) \hat{c}_-(t) \rangle +|\mathcal{V}^{-1}_{12}|^2 \langle \hat{c}_+^{\dg}(t) \hat{c}_+(t)\rangle\nn\\
&&~~~+\mathcal{V}^{-1*}_{12}\mathcal{V}^{-1}_{11} \langle \hat{c}_+^{\dg}(t) \hat{c}_-(t) \rangle +\mathcal{V}^{-1*}_{11}\mathcal{V}^{-1}_{12} \langle \hat{c}_-^{\dg}(t) \hat{c}_+(t)\rangle \big)\kappa_a\nn\\
&&=\frac{16\kappa_a^2\mathcal{I}_{\rm in}}{(4J^2-\gamma^2)^2}\sin^2\big(\frac{t}{4}\sqrt{4J^2-\gamma^2}\big)\Big(2J^2\nn\\&&~~~+(2J^2-\gamma^2)\cos\big(\frac{t}{2}\sqrt{4J^2-\gamma^2}\big)\nn\\&&~~~+\gamma \sqrt{4J^2-\gamma^2}\sin \big(\frac{t}{2}\sqrt{4J^2-\gamma^2}\big)\Big)+\mathcal{I}^{(n)}_{LL}(t)\nn\\
&&=\frac{16\kappa_a^2J^2\mathcal{I}_{\rm in}}{(4J^2-\gamma^2)^2}\big(\cos{\theta}-\cos{(\theta+Jt\sin{\theta})}\big)^2+\mathcal{I}^{(n)}_{LL}(t),\nn\\\label{OILL2}
\eea
where $\cos{\theta}=\gamma/(2J)$, and we have inserted the correlations from Eqs.~\ref{cor1}-\ref{cor4} in deriving the last line. Here, $\mathcal{I}^{(n)}_{LL}(t)$ is the noise contribution to the reflected output intensity in the left optical fiber. We further derive
\bea
&&-\sqrt{\kappa}_a(\langle \hat{a}^{\dg}_{\rm in} \hat{a}(t)\rangle+\langle\hat{a}^{\dg}(t)\hat{a}_{\rm in}\rangle)\nn\\
&&=\frac{8\kappa_aJ\mathcal{I}_{\rm in}}{(4J^2-\gamma^2)}\big(-\frac{\gamma}{2J}+\frac{\gamma}{2J}\cos\big(\frac{t}{2}\sqrt{4J^2-\gamma^2}\big)\nn\\&&~~-\frac{\sqrt{4J^2-\gamma^2}}{2J}\sin\big(\frac{t}{2}\sqrt{4J^2-\gamma^2}\big)\big)\nn\\
&&=-\frac{8\kappa_aJ\mathcal{I}_{\rm in}}{(4J^2-\gamma^2)}\big(\cos{\theta}-\cos{(\theta+Jt\sin{\theta})}\big).\label{OILL3}
\eea
Plugging the contributions from Eqs.~\ref{OILL2},\ref{OILL3} in Eq.~\ref{OILL1}, we get for $\mathcal{I}_{LL}(t)$ by separating the contribution without noise and that due to noise as 
\bea
&&\mathcal{I}_{LL}(t)=\mathcal{I}_{LL}^{(0)}(t)+\mathcal{I}^{(n)}_{LL}(t),\nn\\
&&\mathcal{I}^{(0)}_{LL}(t)=\mathcal{I}_{\rm in}\Big(1-\frac{4\kappa_aJ}{4J^2-\gamma^2}\big(\cos{\theta}-\cos{(\theta+Jt\sin{\theta})}\big)\Big)^2,\nn\\
&&\mathcal{I}^{(n)}_{LL}(t)=\frac{4\kappa_a\gamma}{4J^2-\gamma^2}\Big(J^2t+\gamma\sin^2\big(\frac{t}{2}\sqrt{4J^2-\gamma^2}\:\big)\Big)\nn\\&&~~~~~~~~~~~~+\frac{2\kappa_a\gamma(2J^2-\gamma^2)}{(4J^2-\gamma^2)^{3/2}}\sin\big(t\sqrt{4J^2-\gamma^2}\:\big).\label{OILLn}
\eea
Here, $\mathcal{I}_{LL}^{(0)}(t)$ and $\mathcal{I}^{(n)}_{LL}(t)$ in Eq.~\ref{OILLn}  work both in the unbroken and broken phase of the $\mathcal{PT}$ symmetric system. 

{\it Input from right:} Next we consider an input from the right of the coupled resonators, i.e., $\hat{a}_{\rm in}=0$ and $\hat{b}_{\rm in} \ne 0$. The transmitted output intensity in the left optical fiber of the system is
\bea
&&\mathcal{I}_{RL}(t)=\langle \hat{a}^{\dg}_{\rm out}\hat{a}_{\rm out} \rangle\nn\\
&&=\langle (-\hat{a}^{\dg}_{\rm in}+\sqrt{\kappa_a}\:\hat{a}^{\dg}(t))(-\hat{a}_{\rm in}+\sqrt{\kappa_a}\:\hat{a}(t)) \rangle \nn\\
&&= \kappa_a \langle \hat{a}^{\dg}(t) \hat{a}(t)\rangle \nn\\
&&=\kappa_a\big(|\mathcal{V}^{-1}_{11}|^2 \langle \hat{c}_-^{\dg}(t) \hat{c}_-(t) \rangle +|\mathcal{V}^{-1}_{12}|^2 \langle \hat{c}_+^{\dg}(t) \hat{c}_+(t)\rangle\nn\\
&&+\mathcal{V}^{-1*}_{12}\mathcal{V}^{-1}_{11} \langle \hat{c}_+^{\dg}(t) \hat{c}_-(t) \rangle +\mathcal{V}^{-1*}_{11}\mathcal{V}^{-1}_{12} \langle \hat{c}_-^{\dg}(t) \hat{c}_+(t)\rangle \big).\nn\\\label{OIRL} 
\eea
The correlators in Eq.~\ref{OIRL} are now different from those in Eqs.~\ref{cor1}-\ref{cor4} due to a change in the input light. Nevertheless, we can find them as before, and they are
\bea
\langle \hat{c}_-^{\dg}(t) \hat{c}_-(t) \rangle &=&|\mathcal{V}_{12}|^2\kappa_b \mathcal{I}_{\rm in} \frac{|e^{\lambda_-t}-1|^2}{|\lambda_-|^2}\nn\\&&+|\mathcal{V}_{11}|^2\gamma_G \frac{e^{(\lambda_-+\lambda_-^*)t}-1}{\lambda_-+\lambda_-^*},\label{cor5}\\
\langle \hat{c}_+^{\dg}(t) \hat{c}_+(t) \rangle &=&|\mathcal{V}_{22}|^2\kappa_b \mathcal{I}_{\rm in} \frac{|e^{\lambda_+t}-1|^2}{|\lambda_+|^2}\nn\\&&+|\mathcal{V}_{21}|^2\gamma_G \frac{e^{(\lambda_++\lambda_+^*)t}-1}{\lambda_++\lambda_+^*},\label{cor6}\\
\langle \hat{c}_+^{\dg}(t) \hat{c}_-(t) \rangle &=&\mathcal{V}_{12}\mathcal{V}_{22}^*\kappa_b \mathcal{I}_{\rm in} \frac{(e^{\lambda_-t}-1)(e^{\lambda_+^*t}-1)}{\lambda_-\lambda_+^*}\nn\\&&+\mathcal{V}_{11}\mathcal{V}_{21}^*\gamma_G \frac{e^{(\lambda_-+\lambda_+^*)t}-1}{\lambda_-+\lambda_+^*},\label{cor7}\\
\langle \hat{c}_-^{\dg}(t) \hat{c}_+(t) \rangle &=&\mathcal{V}^*_{12}\mathcal{V}_{22}\kappa_b \mathcal{I}_{\rm in} \frac{(e^{\lambda_-^*t}-1)(e^{\lambda_+t}-1)}{\lambda_-^*\lambda_+}\nn\\&&+\mathcal{V}^*_{11}\mathcal{V}_{21}\gamma_G \frac{e^{(\lambda_-^*+\lambda_+)t}-1}{\lambda_-^*+\lambda_+}.\label{cor8}
\eea
Plugging these correlators in Eq.~\ref{OIRL}, we get
\bea
&&\mathcal{I}_{RL}(t)=\mathcal{I}_{RL}^{(0)}(t)+\mathcal{I}_{RL}^{(n)}(t),\nn\\
&&\mathcal{I}_{RL}^{(0)}(t)=\frac{64J^2\kappa_a\kappa_b\mathcal{I}_{\rm in}}{(4J^2-\gamma^2)^2}\sin^4(\frac{t}{4}\sqrt{4J^2-\gamma^2}),\label{OIRL0}\\
&&\mathcal{I}_{RL}^{(n)}(t)=\mathcal{I}_{LL}^{(n)}(t).
\eea
We thus find $\mathcal{I}_{LR}^{(0)}(t)=\mathcal{I}_{RL}^{(0)}(t)$. The expression of $\mathcal{I}_{RL}^{(0)}(t)$ and $\mathcal{I}_{RL}^{(n)}(t)$ work adequately both in the broken and unbroken phase of the $\mathcal{PT}$ symmetric system. 

\begin{figure}
\includegraphics[width=\linewidth]{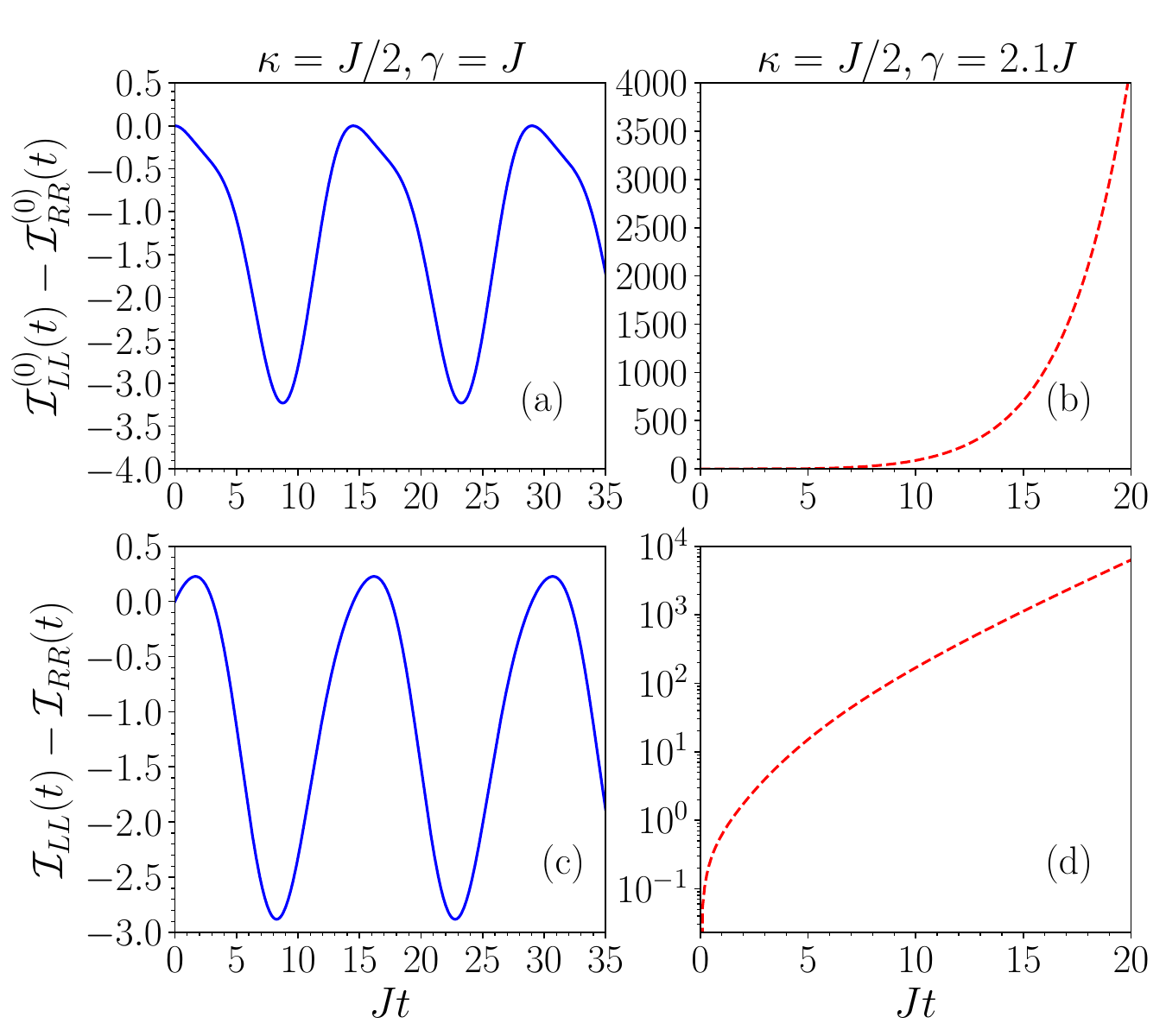}
\caption{Time evolution of asymmetry in left and right reflected output intensity without noise $(\mathcal{I}^{(0)}_{LL}(t)-\mathcal{I}^{(0)}_{RR}(t))$ and with quantum noise $(\mathcal{I}_{LL}(t)-\mathcal{I}_{RR}(t))$ in $\mathcal{PT}$ symmetric coupled resonators in the unbroken (first column) and broken phase (second column). We set $\mathcal{I}_{\rm in}=1$.}
\label{PTsymCRRef}
\end{figure}

 For an input from the right of the coupled resonators, the reflected output intensity in the right optical fiber is
\bea
&&\mathcal{I}_{RR}(t)=\langle \hat{b}^{\dg}_{\rm out}\hat{b}_{\rm out} \rangle\nn\\
&&=\langle (-\hat{b}^{\dg}_{\rm in}+\sqrt{\kappa_b}\:\hat{b}^{\dg}(t))(-\hat{b}_{\rm in}+\sqrt{\kappa_b}\:\hat{b}(t)) \rangle \nn\\
&&= \mathcal{I}_{\rm in}+ \kappa_b \langle \hat{b}^{\dg}(t) \hat{b}(t)\rangle-\sqrt{\kappa}_b(\langle \hat{b}^{\dg}_{\rm in} \hat{b}(t)\rangle+\langle\hat{b}^{\dg}(t)\hat{b}_{\rm in}\rangle).\nn\\\label{OIRR1}
\eea
We apply the correlators from Eqs.~\ref{cor5}-\ref{cor8} in Eq.~\ref{OILR} to get 
\bea
&&\kappa_b\langle \hat{b}^{\dg}(t) \hat{b}(t)\rangle \nn\\
&&=\frac{16\kappa_b^2J^2\mathcal{I}_{\rm in}}{(4J^2-\gamma^2)^2}\big(\cos{\theta}-\cos{(\theta-Jt\sin{\theta})}\big)^2+\mathcal{I}^{(n)}_{LR}(t),\nn \\\label{OIRR2}
\eea
where again $\cos{\theta}=\gamma/(2J)$. We further find
\bea
&&-\sqrt{\kappa}_b(\langle \hat{b}^{\dg}_{\rm in} \hat{b}(t)\rangle+\langle\hat{b}^{\dg}(t)\hat{b}_{\rm in}\rangle)\nn\\
&&=\frac{8\kappa_bJ\mathcal{I}_{\rm in}}{(4J^2-\gamma^2)}\big(\cos{\theta}-\cos{(\theta-Jt\sin{\theta})}\big).\label{OIRR3}
\eea
Adding all the terms in Eq.~\ref{OIRR1}, we derive
\bea
&&\mathcal{I}_{RR}(t)=\mathcal{I}_{RR}^{(0)}(t)+\mathcal{I}^{(n)}_{LR}(t),\nn\\
&&\mathcal{I}^{(0)}_{RR}(t)=\mathcal{I}_{\rm in}\Big(1+\frac{4\kappa_bJ}{4J^2-\gamma^2}\big(\cos{\theta}-\cos{(\theta-Jt\sin{\theta})}\big)\Big)^2,\nn\\
\eea
which shows the reflected output intensities are mostly different at the opposite ends of the system, even without the noise contribution. We find $\mathcal{I}_{LL}^{(0)}(t)=\mathcal{I}_{RR}^{(0)}(t)$ in the unbroken phase at time $t_{m}=4m\pi/\sqrt{4J^2-\gamma^2}$ with $m=0,1,2,3\dots$ as shown in Fig.~\ref{PTsymCRRef}(a). In Figs.~\ref{PTsymCRRef}(a,b), the asymmetry in output reflected intensities without the noise contributions ($\mathcal{I}_{LL}^{(0)}(t)-\mathcal{I}_{RR}^{(0)}(t)$) oscillates with time in the unbroken phase and proliferates with time in the broken phase. The noise contributions to the reflected output intensity are also different for a left and a right incident of light. Thus, the presence of quantum noise further increases the asymmetry in total output reflected intensities $\mathcal{I}_{LL}(t)-\mathcal{I}_{RR}(t)$ in the broken phases, which we show in Fig.~\ref{PTsymCRRef}(d).

\subsection{Steady-state photon transport in coupled resonators}\label{TID}
Next, we will give details of the steady-state photon transport in the coupled resonators. Here, we consider non-zero thermal photon occupancy for general photon frequencies. Since the condition for obtaining steady state of the system requires $\gamma_G \le \gamma$ or $\gamma<\gamma_G<2\gamma$, the matrix $\mathcal{V}_s$ diagonalizing the matrix $\mathcal{M}_r$ is different from $\mathcal{V}$ in Eq.~\ref{UmatCR} for the $\mathcal{PT}$ symmetric case with $\gamma_G=2\gamma$. For the steady-state case, we have
\bea
\mathcal{V}_s=\begin{pmatrix}{\f{i\big(-\gamma_G+\sqrt{\gamma_G^2-16J^2}\:\big)}{4J}} &{1}\\{\f{-i\big(\gamma_G+\sqrt{\gamma_G^2-16J^2}\:\big)}{4J}}&{1}\end{pmatrix},\: \mathcal{V}_s\mathcal{M}_r\mathcal{V}_s^{-1}=\begin{pmatrix}{\Lambda_-} &{0}\\{0}&{\Lambda_+}\end{pmatrix},\nn
\eea
where $\Lambda_{\mp}=(\gamma_G-2\gamma \mp \sqrt{\gamma_G^2-16J^2}\:)/4$ are  eigenvalues of $\mathcal{M}_r$ in Eq.~\ref{evM} for $\delta_a=\delta_b=0$ and $\gamma_a+\kappa_a=\gamma_b+\kappa_b=\gamma$. Thus, we can write Eq.~\ref{ev} in terms of $\hat{c}_{\pm}$ as 
\bea
&&\frac{d}{dt}\hat{c}_{\mp}=\Lambda_{\mp}\hat{c}_{\mp}+ \hat{c}_{{\rm in},\mp}+\hat{f}_{\mp}\:,\begin{pmatrix}{\hat{c}_-}\\{\hat{c}_+}\end{pmatrix}=\mathcal{V}_s\begin{pmatrix}{\hat{a}} \\{\hat{b}}\end{pmatrix},\label{NMs1}\\
&&\begin{pmatrix}{\hat{c}_{{\rm in},-}} \\{\hat{c}_{{\rm in},+}}\end{pmatrix}=\mathcal{V}_s\begin{pmatrix}{\sqrt{\kappa_a}\:\hat{a}_{\rm in}} \\{\sqrt{\kappa_b}\:\hat{b}_{\rm in}}\end{pmatrix},\begin{pmatrix}{\hat{f}_-} \\{\hat{f}_+}\end{pmatrix}=\mathcal{V}_s\begin{pmatrix}{\hat{f}_a} \\{\hat{f}_b}\end{pmatrix}.
\eea
The steady-state solution of these operators in Eq.~\ref{NMs1} at long time for ${\rm Re}[\Lambda_{\pm}]<0$ are
\bea
\hat{c}_{\mp}(t)=-\frac{1}{\Lambda_{\mp}}\hat{c}_{{\rm in},\mp}+\int_{0}^{t}d\tau\:e^{\Lambda_{\mp}(t-\tau)}\hat{f}_{\mp}(\tau).\label{SolNMs}
\eea
For a single-photon input from the left of the coupled resonators, e.g.,  $\hat{a}_{\rm in} \ne 0$ and $\hat{b}_{\rm in}=0$, the steady-state transmitted output intensity in the right optical fiber of the system at a long time is
\bea
&&\mathcal{I}_{LR}=\langle \hat{b}^{\dg}_{\rm out}\hat{b}_{\rm out} \rangle=\kappa_b \lim_{t \to \infty}\langle \hat{b}^{\dg}(t) \hat{b}(t)\rangle \nn\\
&&= \lim_{t \to \infty}\kappa_b\big(|\mathcal{V}^{-1}_{s22}|^2 \langle \hat{c}_+^{\dg}(t) \hat{c}_+(t) \rangle +|\mathcal{V}^{-1}_{s21}|^2 \langle \hat{c}_-^{\dg}(t) \hat{c}_-(t)\rangle\nn\\
&&+\mathcal{V}^{-1*}_{s22}\mathcal{V}^{-1}_{s21} \langle \hat{c}_+^{\dg}(t) \hat{c}_-(t) \rangle +\mathcal{V}^{-1*}_{s21}\mathcal{V}^{-1}_{s22} \langle \hat{c}_-^{\dg}(t) \hat{c}_+(t)\rangle \big),\nn\\\label{OILRs}
\eea
where $\mathcal{V}^{-1}_{sij}$ are elements of $\mathcal{V}^{-1}_s$. The steady-state correlators in Eq.~\ref{OILRs} can be found using the formal solutions of the operators in Eq.~\ref{SolNMs} and the noise properties in Eq.~\ref{fd2}. We find
\bea
&&\langle \hat{c}_-^{\dg}(t) \hat{c}_-(t) \rangle =\frac{\kappa_a \mathcal{I}_{\rm in}|\mathcal{V}_{s11}|^2}{|\Lambda_-|^2}\nn\\&&~~~~~-\frac{\big(|\mathcal{V}_{s11}|^2(\gamma_G+\gamma_a n_{\rm th})+|\mathcal{V}_{s12}|^2\gamma_b n_{\rm th}\big)}{\Lambda_-+\Lambda_-^*},\label{cor1s}\\
&&\langle \hat{c}_+^{\dg}(t) \hat{c}_+(t) \rangle =\frac{\kappa_a \mathcal{I}_{\rm in}|\mathcal{V}_{s21}|^2}{|\Lambda_+|^2}\nn\\&&~~~~~-\frac{\big(|\mathcal{V}_{s21}|^2(\gamma_G+\gamma_a n_{\rm th})+|\mathcal{V}_{s22}|^2\gamma_b n_{\rm th}\big)}{\Lambda_++\Lambda_+^*},\label{cor2s}\\
&&\langle \hat{c}_+^{\dg}(t) \hat{c}_-(t) \rangle =\frac{\kappa_a \mathcal{I}_{\rm in}\mathcal{V}_{s11}\mathcal{V}^*_{s21}}{\Lambda_-\Lambda_+^*}\nn\\&&~-\frac{\big(\mathcal{V}_{s11}\mathcal{V}_{s21}^*(\gamma_G+\gamma_a n_{\rm th})+\mathcal{V}_{s12}\mathcal{V}_{s22}^*\gamma_b n_{\rm th}\big)}{\Lambda_-+\Lambda_+^*},\label{cor3s}\\
&&\langle \hat{c}_-^{\dg}(t) \hat{c}_+(t) \rangle =\frac{\kappa_a \mathcal{I}_{\rm in}\mathcal{V}_{s11}^*\mathcal{V}_{s21}}{\Lambda_-^*\Lambda_+}\nn\\&&~-\frac{\big(\mathcal{V}_{s11}^*\mathcal{V}_{s21}(\gamma_G+\gamma_a n_{\rm th})+\mathcal{V}_{s12}^*\mathcal{V}_{s22}\gamma_b n_{\rm th}\big)}{\Lambda_-^*+\Lambda_+}.\label{cor4s}
\eea
We insert these correlators from Eqs.~\ref{cor1s}-\ref{cor4s} in Eq.~\ref{OILRs} and simplify it by using the explicit forms of $\Lambda_{\mp},\mathcal{V}_{sij}$, and $\mathcal{V}^{-1}_{sij}$. The steady-state transmitted output intensity due to the contribution without and with the quantum noise are 
\bea
\mathcal{I}_{LR}&=&\mathcal{I}_{LR}^{(0)}+\mathcal{I}_{LR}^{(n)},\nn\\
\mathcal{I}_{LR}^{(0)}&=&\frac{16J^2\kappa_a\kappa_b\mathcal{I}_{\rm in}}{(4J^2+\gamma(\gamma-\gamma_G))^2},\label{IOLR0s}\\
\mathcal{I}_{LR}^{(n)}&=&\frac{4J^2\kappa_b(\gamma_G+\gamma_a n_{\rm th})}{(2\gamma-\gamma_G)(4J^2+\gamma(\gamma-\gamma_G))}\nn\\&+&\frac{\kappa_b\gamma_bn_{\rm th}(4J^2+(\gamma-\gamma_G)(2\gamma-\gamma_G))}{(2\gamma-\gamma_G)(4J^2+\gamma(\gamma-\gamma_G))}, \label{IOLRns}
\eea
which matches to $\mathcal{I}_{LR}$ in Eq.~\ref{transLRsp} in the limit of $n_{\rm th} \to 0$. A similar calculation for a single-photon input from the right of the coupled resonators ($\hat{a}_{\rm in}=0,~\hat{b}_{\rm in}\ne 0$) gives the following for steady-state transmitted output intensity in the left optical fiber of the system at a long time:
\bea
&&\mathcal{I}_{RL}=\langle \hat{a}^{\dg}_{\rm out}\hat{a}_{\rm out} \rangle=\kappa_a \lim_{t\to \infty}\langle \hat{a}^{\dg}(t) \hat{a}(t)\rangle \nn\\
&&=\lim_{t \to \infty}\kappa_a\big(|\mathcal{V}^{-1}_{s11}|^2 \langle \hat{c}_-^{\dg}(t) \hat{c}_-(t) \rangle +|\mathcal{V}^{-1}_{s12}|^2 \langle \hat{c}_+^{\dg}(t) \hat{c}_+(t)\rangle\nn\\
&&+\mathcal{V}^{-1*}_{s12}\mathcal{V}^{-1}_{s11} \langle \hat{c}_+^{\dg}(t) \hat{c}_-(t) \rangle +\mathcal{V}^{-1*}_{s11}\mathcal{V}^{-1}_{s12} \langle \hat{c}_-^{\dg}(t) \hat{c}_+(t)\rangle \big).\label{OIRLs} 
\eea
We calculate the steady-state correlators in Eq.~\ref{OIRLs} for input from the right and plug them in Eq.~\ref{OIRLs} to find two parts of $\mathcal{I}_{RL}$ as
\bea
\mathcal{I}_{RL}&=&\mathcal{I}_{RL}^{(0)}+\mathcal{I}_{RL}^{(n)},\nn\\
\mathcal{I}_{RL}^{(0)}&=&\mathcal{I}_{LR}^{(0)},\\
\mathcal{I}_{RL}^{(n)}&=&\frac{\kappa_a(4J^2+\gamma(2\gamma-\gamma_G))(\gamma_G+\gamma_a n_{\rm th})}{(2\gamma-\gamma_G)(4J^2+\gamma(\gamma-\gamma_G))}\nn\\&+&\frac{4J^2\kappa_a\gamma_bn_{\rm th}}{(2\gamma-\gamma_G)(4J^2+\gamma(\gamma-\gamma_G))}, \label{IORLns}
\eea
which again matches with $\mathcal{I}_{RL}$ in Eq.~\ref{transRLsp} when $n_{\rm th}\to 0$.

\section{Two waveguides coupled in parallel}\label{WG}
In this section, we give details of the calculation to find outgoing transmitted intensities from two finite-length waveguides evanescently coupled in parallel. The calculations are related to the previous case of two direct coupled resonators. Mainly, we show below that the transmitted output intensity of an incident photon from $z = 0$ of the waveguide $A~(B)$ to $z = l$ of the waveguide $B~(A)$ and $A~(B)$ are related, respectively, to the transmitted and reflected output intensity of an incident photon from the left (right) of the coupled resonators. Nevertheless, there are specific differences in the magnitude of output intensities without the quantum noise between the two models due to differences in injecting photons in the two systems. We directly populate the photon modes in the waveguides as given in Eq.~\ref{qf1}.

We find the spatial evolution of quantum light fields in Eq.~\ref{qf1} by introducing a matrix made of left eigenvectors of $\mathcal{M}_w$ with $\mathcal{PT}$ symmetry for  $\gamma_G=2\gamma$. Since $\mathcal{M}_w$ is identical to $\mathcal{PT}$ symmetric $\mathcal{M}_r$ in Eq.~\ref{UmatCR}, the eigenvalues $(\lambda_{\pm})$ and eigenvectors of $\mathcal{M}_w$ are the same as $\mathcal{M}_r$. Therefore, the diagonalizing matrix for this case is also $\mathcal{V}$ of Eq.~\ref{UmatCR},
which leads to the following uncoupled equations:
\bea
\frac{d}{dz}\hat{c}_{\mp}(z)&=&\lambda_{\mp}\hat{c}_{\mp}(z)+\hat{f}_{\mp}(z),\label{NMw1}\\{\rm with}~\begin{pmatrix}{\hat{c}_-} \\{\hat{c}_+}\end{pmatrix}&=&\mathcal{V}\begin{pmatrix}{\hat{a}} \\{\hat{b}}\end{pmatrix},~{\rm and}~\begin{pmatrix}{\hat{f}_-} \\{\hat{f}_+}\end{pmatrix}=\mathcal{V}\begin{pmatrix}{\hat{f}_a} \\{\hat{f}_b}\end{pmatrix}.
\eea
The formal solution for spatial evolution of the operators in Eq.~\ref{NMw1} from the left $(z=0)$ to right $(z=l)$ of the waveguides are
\bea
\hat{c}_{\mp}(l)=e^{\lambda_{\mp}l}\hat{c}_{\mp}(0)+\int_{0}^{l}dz\:e^{\lambda_{\mp}(l-z)}\hat{f}_{\mp}(z),\label{SolNMw}
\eea
where $\hat{c}_{\mp}(0)$ are the initial population of these operators at $z=0$.

{\it Input in waveguide A:} For a single-photon input in the waveguide $A$ of the coupled waveguides, i.e., $\hat{a}(0) \ne 0$ and $\hat{b}(0)=0$, the transmitted output intensity at $z=l$ of the waveguide $A$ of the system is
\bea
&&\mathcal{I}_{AA}(l)=\langle 1,0|\hat{a}^{\dg}(l)\hat{a}(l)|1,0 \rangle=\mathcal{I}_{AA}^{(0)}(l)+\mathcal{I}_{A}^{(n)}(l)\nn\\
&&=\big(|\mathcal{V}^{-1}_{12}|^2 \langle \hat{c}_+^{\dg}(l) \hat{c}_+(l) \rangle +|\mathcal{V}^{-1}_{11}|^2 \langle \hat{c}_-^{\dg}(l) \hat{c}_-(l)\rangle\nn\\
&&+\mathcal{V}^{-1*}_{12}\mathcal{V}^{-1}_{11} \langle \hat{c}_+^{\dg}(l) \hat{c}_-(l) \rangle +\mathcal{V}^{-1*}_{11}\mathcal{V}^{-1}_{12} \langle \hat{c}_-^{\dg}(l) \hat{c}_+(l)\rangle \big),\label{OIAA}
\eea
which is similar to the part of reflected output intensity (first two lines of Eq.~\ref{OILL2}) for an incident photon from the left of coupled resonators. We find the correlators in Eq.~\ref{OIAA} employing the formal solutions of the operators in Eq.~\ref{SolNMw} and the correlation properties of quantum noises. We further apply $\langle 1,0|\hat{a}^{\dg}(0)\hat{a}(0)|1,0\rangle=1$ to get the following:
\bea
\langle \hat{c}_-^{\dg}(l) \hat{c}_-(l) \rangle &=&|\mathcal{V}_{11}|^2 \big(e^{(\lambda_-+\lambda_-^*)l}+\gamma_G \frac{e^{(\lambda_-+\lambda_-^*)l}-1}{\lambda_-+\lambda_-^*}\big),\nn\\\label{cor1w}\\
\langle \hat{c}_+^{\dg}(l) \hat{c}_+(l) \rangle &=&|\mathcal{V}_{21}|^2 \big(e^{(\lambda_++\lambda_+^*)l}+\gamma_G \frac{e^{(\lambda_++\lambda_+^*)l}-1}{\lambda_++\lambda_+^*}\big),\nn\\\label{cor2w}\\
\langle \hat{c}_+^{\dg}(l) \hat{c}_-(l) \rangle &=&\mathcal{V}_{11}\mathcal{V}_{21}^*\Big(e^{(\lambda_-+\lambda_+^*)l}+\gamma_G \frac{e^{(\lambda_-+\lambda_+^*)l}-1}{\lambda_-+\lambda_+^*}\Big),\nn\\\label{cor3w}\\
\langle \hat{c}_-^{\dg}(l) \hat{c}_+(l) \rangle &=&\mathcal{V}^*_{11}\mathcal{V}_{21}\Big(e^{(\lambda_-^*+\lambda_+)l}+\gamma_G \frac{e^{(\lambda_-^*+\lambda_+)l}-1}{\lambda_-^*+\lambda_+}\Big).\nn\\\label{cor4w}
\eea
We thus get for the noise-free contribution, $\mathcal{I}_{AA}^{(0)}$, and the contribution due to noise, $\mathcal{I}_{A}^{(n)}$, by inserting Eqs.~\ref{cor1w}-\ref{cor4w} in Eq.~\ref{OIAA}:
\bea
\mathcal{I}_{AA}^{(0)}(l)&=&\Big(\cos\big(\frac{l}{2}\sqrt{4J^2-\gamma^2}\:\big)\nn\\&+&\frac{\gamma}{\sqrt{4J^2-\gamma^2}}\sin\big(\frac{l}{2}\sqrt{4J^2-\gamma^2}\:\big) \Big)^2, \\
\mathcal{I}_{A}^{(n)}(l)&=&\frac{4\gamma}{4J^2-\gamma^2}\Big(J^2l+\gamma\sin^2\big(\frac{l}{2}\sqrt{4J^2-\gamma^2}\:\big)\Big)\nn\\&+&\frac{2\gamma(2J^2-\gamma^2)}{(4J^2-\gamma^2)^{3/2}}\sin\big(l\sqrt{4J^2-\gamma^2}\:\big).\label{OIAAn}
\eea
Here, $\mathcal{I}_{A}^{(n)}(l)$ in Eq.~\ref{OIAAn} is the same as the noise contribution ($\mathcal{I}_{LL}^{(n)}(t)$ in Eq.~\ref{OILLn}) to the reflected output intensity of an incident single photon from the left of the coupled resonators when we set $\kappa_a=1$ and replace $t$ by $l$. However, $\mathcal{I}_{AA}^{(0)}(l)$ is not precisely similar to $\mathcal{I}_{LL}^{(0)}(t)$ due to differences in injecting photons in two coupled systems.

The transmitted output intensity at $z=l$ of the waveguide $B$ of the system for a single-photon input in the waveguide $A$ is
\bea
&&\mathcal{I}_{AB}(l)=\langle 1,0|\hat{b}^{\dg}(l)\hat{b}(l)|1,0 \rangle=\mathcal{I}_{AB}^{(0)}(l)+\mathcal{I}_{B}^{(n)}(l)\nn\\
&&=\big(|\mathcal{V}^{-1}_{22}|^2 \langle \hat{c}_+^{\dg}(l) \hat{c}_+(l) \rangle +|\mathcal{V}^{-1}_{21}|^2 \langle \hat{c}_-^{\dg}(l) \hat{c}_-(l)\rangle\nn\\
&&+\mathcal{V}^{-1*}_{22}\mathcal{V}^{-1}_{21} \langle \hat{c}_+^{\dg}(l) \hat{c}_-(l) \rangle +\mathcal{V}^{-1*}_{21}\mathcal{V}^{-1}_{22} \langle \hat{c}_-^{\dg}(l) \hat{c}_+(l)\rangle \big),\nn\\\label{OIAB}
\eea
which is similar to the transmitted output intensity in Eq.~\ref{OILR} for an incident photon from the left of coupled resonators. We insert the correlators in Eqs.~\ref{cor1w}-\ref{cor4w} in Eq.~\ref{OIAB} to find the noise-free and noise contributions to $\mathcal{I}_{AB}(l)$, respectively, as
\bea
\mathcal{I}_{AB}^{(0)}(l)&=& \frac{4J^2}{4J^2-\gamma^2}\sin^2\big(\f{l}{2}\sqrt{4J^2-\gamma^2}\: \big),\\
\mathcal{I}_{B}^{(n)}(l)&=&\frac{4\gamma J^2}{4J^2-\gamma^2}\big(l-\frac{1}{\sqrt{4J^2-\gamma^2}}\sin\big(l\sqrt{4J^2-\gamma^2}\:\big)\big).\nn\\\label{OIABn}
\eea
Again, $\mathcal{I}_{B}^{(n)}(l)$ in Eq.~\ref{OIABn} is the same as the noise contribution ($\mathcal{I}_{LR}^{(n)}(t)$ in Eq.~\ref{IOLRn}) to the transmitted output intensity of a single-photon input from the left of the coupled resonators when we set $\kappa_b=1$ and replace $t$ by $l$. 

{\it Input in waveguide B:} We next consider a single-photon input in the waveguide $B$ of the coupled waveguides, i.e., $\hat{a}(0)=0$ and $\hat{b}(0)\ne 0$. The transmitted output intensity at $z=l$ of the waveguide $B$ of the system is
\bea
&&\mathcal{I}_{BB}(l)=\langle 0,1|\hat{b}^{\dg}(l)\hat{b}(l)|0,1 \rangle \nn\\
&&=\big(|\mathcal{V}^{-1}_{22}|^2 \langle \hat{c}_+^{\dg}(l) \hat{c}_+(l) \rangle +|\mathcal{V}^{-1}_{21}|^2 \langle \hat{c}_-^{\dg}(l) \hat{c}_-(l)\rangle\nn\\
&&+\mathcal{V}^{-1*}_{22}\mathcal{V}^{-1}_{21} \langle \hat{c}_+^{\dg}(l) \hat{c}_-(l) \rangle +\mathcal{V}^{-1*}_{21}\mathcal{V}^{-1}_{22} \langle \hat{c}_-^{\dg}(l) \hat{c}_+(l)\rangle \big),\nn\\\label{OIBB}
\eea
which is similar to the part of reflected output intensity for an incident photon from the right of coupled resonators. We calculate the correlators in Eq.~\ref{OIBB} for the incident photon in waveguide $B$ and plug them in Eq.~\ref{OIBB} to find
\bea
\mathcal{I}_{BB}(l)&=&\mathcal{I}_{BB}^{(0)}(l)+\mathcal{I}_{B}^{(n)}(l),\nn\\
\mathcal{I}_{BB}^{(0)}(l)&=&\Big(\cos\big(\frac{l}{2}\sqrt{4J^2-\gamma^2}\:\big)\nn\\&-&\frac{\gamma}{\sqrt{4J^2-\gamma^2}}\sin\big(\frac{l}{2}\sqrt{4J^2-\gamma^2}\:\big) \Big)^2.
\eea
Since the contribution to the correlators from the quantum noises (e.g., the second parts of Eqs.~\ref{cor1w}-\ref{cor4w}) does not change when we switch the waveguide of an incident photon, the noise contributions to the output intensities are thus related for two different initial conditions. A similar calculation for the transmitted output intensity at $z=l$ of the waveguide $A$ of the system is
\bea
&&\mathcal{I}_{BA}(l)=\langle 0,1|\hat{a}^{\dg}(l)\hat{a}(l)|0,1 \rangle=\mathcal{I}_{BA}^{(0)}(l)+\mathcal{I}_{A}^{(n)}(l),\nn\\
&&\mathcal{I}_{BA}^{(0)}(l)=\mathcal{I}_{AB}^{(0)}(l). \label{recWG}
\eea

\begin{figure}[H]
\includegraphics[width=\linewidth]{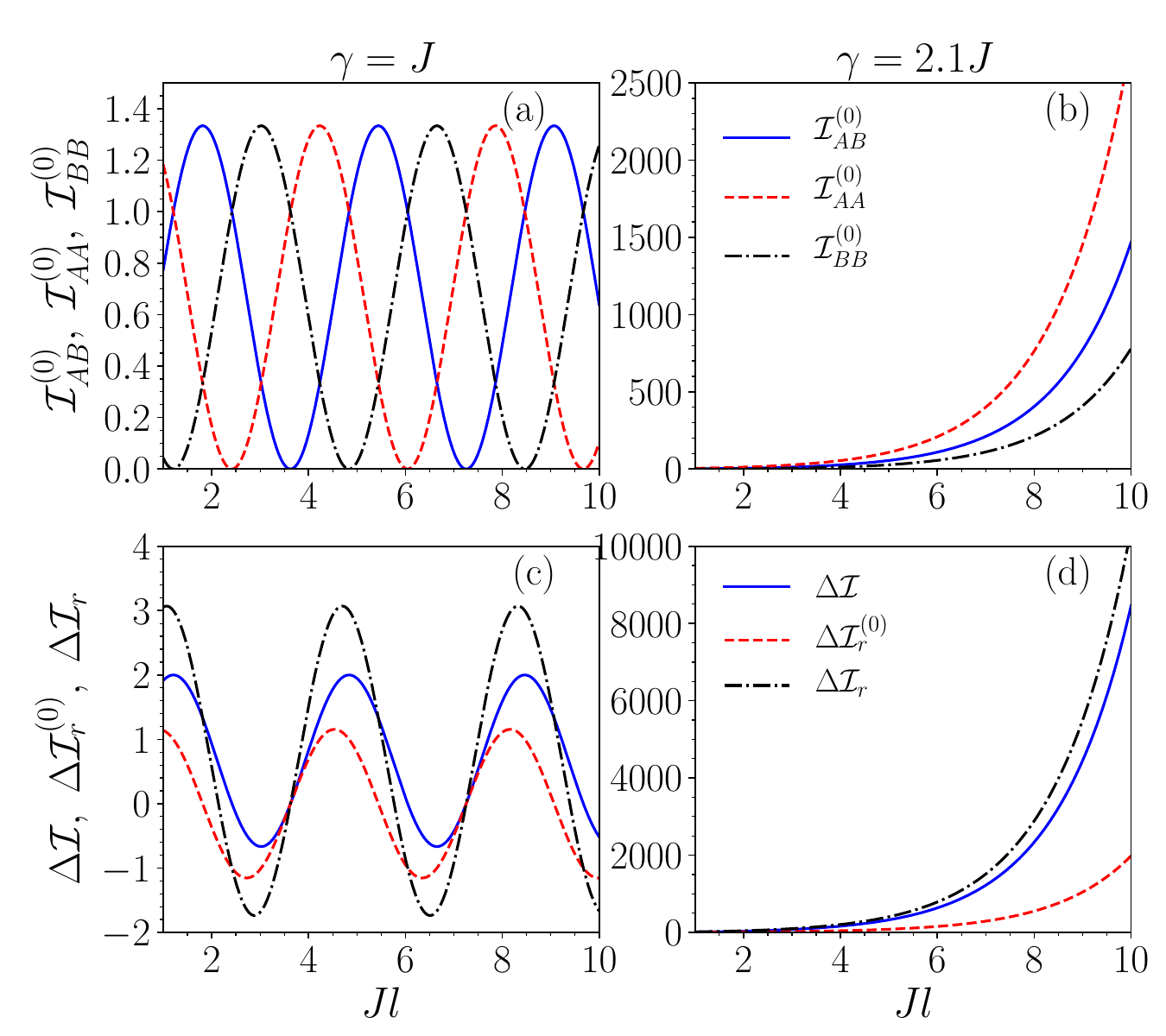}
\caption{Dependence of transmitted output intensity without noise $(\mathcal{I}^{(0)}_{AB},\mathcal{I}^{(0)}_{AA},\mathcal{I}^{(0)}_{BB})$ and nonreciprocity due to quantum noise $(\Delta \mathcal{I}=\mathcal{I}_{BA}-\mathcal{I}_{AB})$ on scaled length $(Jl)$ of the waveguides in $\mathcal{PT}$ symmetry unbroken (first column) and broken phase (second column) of the coupled waveguides. We also plot $\Delta \mathcal{I}_r^{(0)}=\mathcal{I}^{(0)}_{AA}-\mathcal{I}^{(0)}_{BB}$ and $\Delta \mathcal{I}_r=\mathcal{I}_{AA}-\mathcal{I}_{BB}$ with $Jl$ in (c,d).}
\label{PTsymCW}
\end{figure}

In the absence of quantum noises, the transmission of a single photon from waveguide $A$ to waveguide $B$ is the same as the transmission of a single photon from waveguide $B$ to waveguide $A$ both in the $\mathcal{PT}$ symmetry unbroken and broken phase of the coupled system (i.e., Eq.~\ref{recWG}). Nevertheless, the transmission $(\mathcal{I}_{AA}^{(0)}(l))$ of a single photon in the waveguide $A$ with an active gain is higher than that $(\mathcal{I}_{BB}^{(0)}(l))$ in the waveguide $B$ with the only loss in the broken phase even in the absence of quantum noises as shown in Fig.~\ref{PTsymCW}(b). In Figs.~\ref{PTsymCW}(a,b), we compare $\mathcal{I}_{AA}^{(0)}(l),\mathcal{I}_{BB}^{(0)}(l),\mathcal{I}_{AB}^{(0)}(l)$ with an increasing waveguide length $l$ in the $\mathcal{PT}$ symmetry unbroken $(\gamma<2J)$ and broken $(\gamma>2J)$ phase of the coupled system. While these output intensities oscillate with $l$ in the unbroken phase, they increase rapidly with $l$ in the broken phase. Quantum noise leads to a nonreciprocity in the output intensities between two waveguides, i.e., $\mathcal{I}_{BA}(l) \ne \mathcal{I}_{AB}(l)$. The nonreciprocity $\Delta \mathcal{I}(l)=\mathcal{I}_{BA}(l)-\mathcal{I}_{AB}(l)$ oscillates with $l$ in the unbroken phase, and it diverges with $l$ in the broken phase as shown in Figs.~\ref{PTsymCW}(c,d). We also plot $\delta \mathcal{I}^{(0)}_r(l)=\mathcal{I}_{AA}^{(0)}(l)-\mathcal{I}_{BB}^{(0)}(l)$ and $\delta \mathcal{I}_r(l)=\mathcal{I}_{AA}(l)-\mathcal{I}_{BB}(l)$ with $l$ in Figs.~\ref{PTsymCW}(c,d). Fig.~\ref{PTsymCW}(d) displays that the quantum noise greatly enhances the asymmetry in transmission in the same waveguide between two different injections of photons in the broken phase, i.e., $\delta \mathcal{I}_r(l) \gg \delta \mathcal{I}^{(0)}_r(l)$. 

 \bibliography{references}
\end{document}